\documentclass[12pt]{iopart}
\usepackage{graphicx}

\begin{document}

\title[Dynamic interactions between vortex species in $Bi_{2}Sr_{2}CaCu_{2}O_{8}$]
{Mapping the dynamic interactions between vortex species in highly anisotropic superconductors}

\author{M Tesei$^1$, G K Perkins$^1$, A D Caplin$^1$, L F Cohen$^1$ and T Tamegai$^2$}

\address{$^1$ The Blackett Laboratory, Imperial College, London SW7 2BZ, UK}
\address{$^2$ Department of Applied Physics, The University of Tokyo, Hongo, Bunkyo-ku, Tokyo 113-8656, Japan}

\ead{m.tesei@imperial.ac.uk}

\begin{abstract}
Here we use highly sensitive magnetisation measurements performed using a Hall probe sensor on single crystals of
highly anisotropic high temperature superconductors $Bi_{2}Sr_{2}CaCu_{2}O_{8}$ to study the dynamic interactions
between the two species of vortices that exist in such superconductors. We observe a remarkable and clearly delineated
high temperature regime that mirrors the underlying vortex phase diagram. Our results map out the parameter space over
which these dynamic interaction processes can be used to create vortex ratchets, pumps and other fluxonic devices.
\end{abstract}

\pacs{74.25.Qt,74.72.Hs} 


\section{Introduction}

The controlled manipulation of flux quanta (known as vortices or fluxons) within superconductors has recently been
proposed as the basis for an entirely new class of devices \cite{savelev-fluxonics devices}. Collectively known as
fluxonics, applications are far-reaching. Examples include rapid single flux quantum (RSFQ) logic devices for high
speed computing and quantum computing, the control of many-body flux distributions in flux optics devices (flux
lenses), as well as flux ratchets and pumps that could be used to disperse/concentrate flux in fluxonic nanodevices.
There has been significant progress theoretically, particularly by molecular dynamics computer simulations
\cite{Olson-pumps,Wambaugh-pumps}. It has been hypothesised \cite{savelev-fluxonics devices} that in certain
anisotropic layered superconductors such as $Bi_{2}Sr_{2}CaCu_{2}O_{8}$ the dragging interactions between two separate
families of fluxons, Josephson vortices (JVs) and stacks of pancake vortices (PVs), that are known to co-exist in such
materials \cite{review,Grigorenko-chain state,Koshelev-tilted field} could be used to manipulate the flux, thereby
circumventing the need for complex nanofabrication in the devices. Experimental progress has been slow, although flux
imaging experiments have shown that there is a significant static interaction between the two systems
\cite{Grigorenko-chain state}, and recently there has been encouraging evidence of flux ratcheting effects closely
resembling predictions \cite{Cole-ratchet}. However, the nature of the dynamic interactions is poorly understood and
little studied \cite{Perkins-dynamics interaction}. The present paper demonstrates that the interactions between PVs
and JVs are complex and show several different regimes of behaviour due in part to the complexity of the underlying
vortex phase diagram in this material. Our experimental results can be used to set out the anticipated parameter ranges
(field, temperature, frequency) for optimum fluxonics device performance.

\section{Experimental technique}

$Bi_{2}Sr_{2}CaCu_{2}O_{8}$ single crystals ($T_{c}\cong91$K determined by magnetisation measurement with temperature,
size$\approx$1x1mm$^{2}$) were grown by floating-zone technique, for more details see ref \cite{Tamegai-samples}. The
crystals are exposed to a DC perpendicular (to the crystallographic ab-planes) magnetic field $H_{z}$ generated by a
solenoid coil that introduces PVs into the crystal. The perpendicular magnetisation, associated with PVs, is monitored
using an InSb Hall sensor (Te-doped 2$\mu$m thick InSb on GaAs wafer) of active area 100x100$\mu$m$^{2}$ and positioned
above the center of the sample.

\subsection{Lensing}

For DC dragging interaction experiments we use a DC magnetic field $H_{ab}$ that is aligned parallel to the ab planes
of the crystal. For these experiments the sensor is driven by an AC current of 20mA (rms) at 2kHz, with a noise
threshold of 3mG/$\sqrt{\mathrm{Hz}}$. The background magnetic induction $B_{OFF}$ is measured 1.5mm above the sample,
where the sample-induced magnetic screening is reduced to less than 1\%. Although the $H_{ab}$ field and the crystal
plane are aligned to within 0.4 degree, there is unavoidable misalignment between $H_{ab}$ and the sensor plane and
this results in a linearly-dependent background which is subtracted after data has been recorded at fixed height
20$\mu$m above the sample surface, $B_{ON}$, resulting in the sample self-induction $\Delta B=B_{ON}-B_{OFF}$. For
these experiments the sample is cooled at fixed $H_{z}$, creating a specific PV density, and the magnetic induction is
then probed as described above whilst sweeping $H_{ab}$ at a rate 2Oe/sec.

Before we introduce the technique used to probe the JV-PV interactions it is useful to first show the results of the DC
dragging interactions or lensing. Lensing refers to the process whereby JVs produced by $H_{ab}$ drag the pre-existing
PVs (produced by field cooling in applied $H_{z}$), across the sample to the sample centre and increasing the PV
density. The key signature of lensing is a V-shaped local induction vs field loop, as shown in the inset of
Fig.\ref{fig1}. Note that symmetrisation of the raw data \cite{Cole-lensing} will yield precise mirror image curves in
opposite quadrants, but here we show data without prior symmetrisation. Our experimental results shown in the inset of
Fig.\ref{fig1} confirm the previously-reported lensing interaction \cite{Cole-lensing,Tamegai-lensing}. It is the
symmetry of the curve (with respect to the sign of the ab plane field) that naturally suggests the ultra sensitive
second-harmonic detection technique as described in the following.

\subsection{Dynamic interactions}

The second-harmonic detection technique involves exciting the sample with a superimposed AC (sinusoidal) magnetic field
$H^{AC}_{ab}$ aligned with the sample plane, of maximum amplitude 2Oe and frequency $\omega/2\pi$, which drives JVs
through the sample. The technique relies on the fact that the dragging interactions are symmetric with respect to the
sign of the ab plane field as discussed above. It then follows directly that the dragging forces do not depend on JV
vorticity and the time dependence of the driving force is given by the absolute value of the sinusoidal excitation,
$|H^{AC}_{ab}|$. Simply speaking this means that moving JVs in, or out, of the sample produces the same effect on the
pancake lattice than moving anti-JVs in, or out respectively. Such a driving force $\sim |H^{AC}_{ab}|$ is similar to
an excitation field $H^{AC}_{ab}$ at twice the frequency (2$\omega$) with a phase lag of $\pi/2$ \cite{Perkins-dynamics
interaction}. Thus the output from the Hall sensor (which is driven now by a DC current), is locked on the second
harmonic (2$\omega$) response $B''$ at zero in-plane DC field ($H_{ab}$=0) and we use the quadrature component at
$2\omega$, $B''_{2}$, to measure the JV-PV interactions. $B''_{2}$ is monitored while sweeping the temperature or the
excitation frequency $\omega/2\pi$. When measuring the response as a function of $H_{z}$, the sample is zero-field
cooled and then $H_{z}$ is swept from zero (residual field) to positive and negative fields. Since we measure the
second harmonic response, residual pick-up associated with the first harmonic is rejected and the noise threshold is
two orders of magnitude smaller, of the order of 10$\mu$G/$\sqrt{\mathrm{Hz}}$.

\section{Results and discussion}

\begin{figure}
\begin{center}
\includegraphics{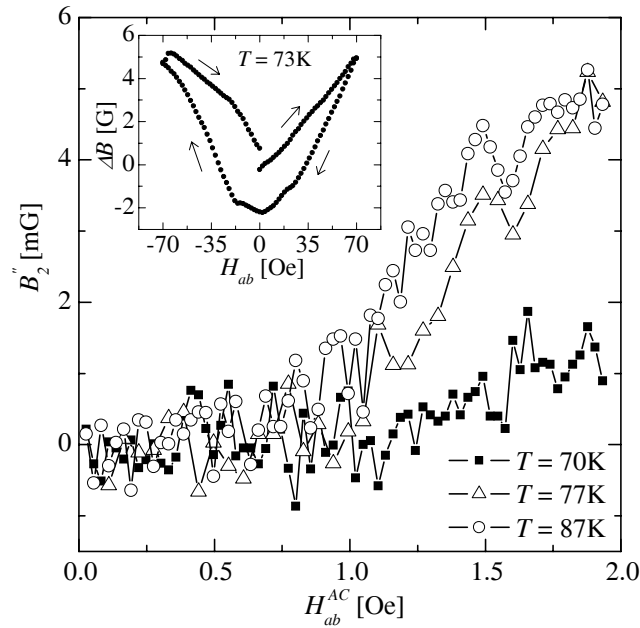}
\end{center}
\caption{\label{fig1} Quadrature component of the 2nd harmonic signal $B''_{2}$ as a function of in-plane AC field
amplitude at $H_{z}$=8Oe, $\omega/2\pi=120$Hz and for three different temperatures corresponding to three different
regimes. Inset: Sample self-induction $\Delta B=B_{ON}-B_{OFF}$ isotherm measured as a function of $H_{ab}$ at
$H_{z}$=10Oe and temperature $T=73$K. Arrows indicate the change in $H_{ab}$. The curve exhibits a total signal change
corresponding to $\sim~70\%$ of the initial PV density.}
\end{figure}

To demonstrate the sensitivity of the 2nd harmonic technique we first show that it is able to capture the onset of the
formation of the crossing lattice, i.e. the formation of separate JVs and PVs states. This onset field has only been
measured previously by Hall probe microscopy \cite{Grigorenko-chain state}, and it is consistent with our measurements.
Fig.\ref{fig1} shows the variation of $B''_{2}$ with amplitude of $H^{AC}_{ab}$ for several temperatures. As can be
seen in the figure, there is a temperature-independent (within the measurement noise) minimum amplitude
$H^{AC}_{ab}\cong0.7$Oe to drive the PVs (for $H_{z}$=8Oe) with the amplitude of the Josephson oscillations. This is
the threshold where the separate lattice states are formed. In our 2nd harmonic technique we do not want to overly
disturb the system and consequently we use a relatively small $H^{AC}_{ab}$ field of 2Oe amplitude for most of the
experiments described here to act as a minor perturbation only. Having set up the parameters for the experiment we are
able to characterise the vortex interactions as a function of temperature, pancake density and AC frequency.

\begin{figure}
\begin{center}
\includegraphics{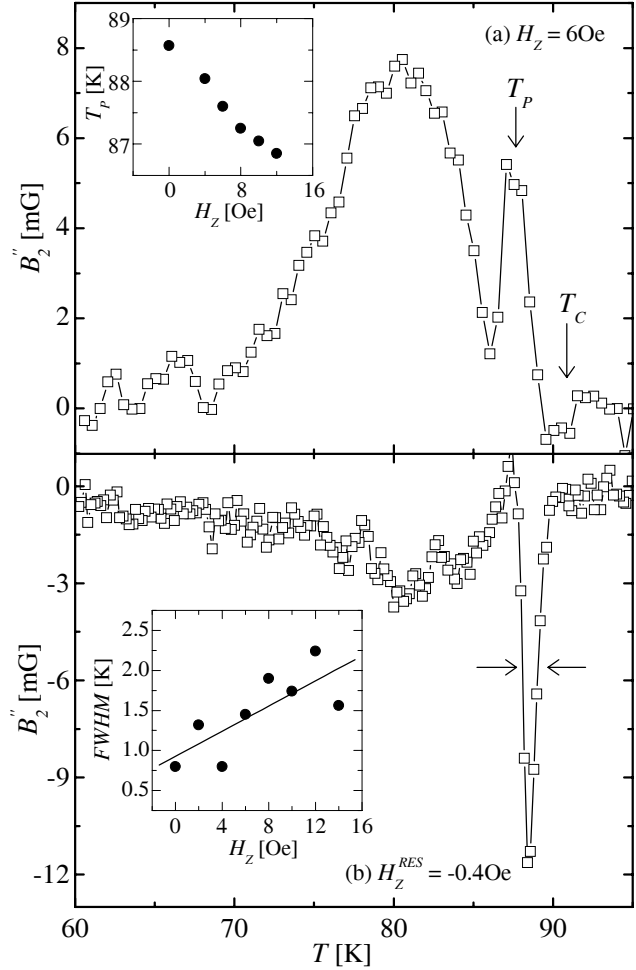}
\end{center}
\caption{\label{fig2} Temperature dependence of 2nd harmonic quadrature component $B''_{2}$ measured at (a) $H_{z}=6$Oe
and at (b) residual negative field $H_{z}^{RES}$=-0.4Oe with excitation amplitude $H^{AC}_{ab}=2$Oe and frequency
$\omega/2\pi=120$Hz. Upper inset: Effect of $H_{z}$ on peak temperature ($T_{p}$). Lower inset: Evolution of the
full-width at half-maximum (FWHM), illustrated with horizontal arrows, as a function of $H_{z}$.}
\end{figure}

Fig.\ref{fig2} shows the evolution of $B''_{2}$ whilst field cooling the sample from above $T_{c}$. We use two values
of the perpendicular magnetic field, $H_{z}$=6Oe and the residual ambient field $H^{RES}_{z}$=-0.4Oe. We observe two
quite distinct regimes of behaviour. A quite remarkable sharp peak close to $T_{c}$ and a broad bump that extends
approximately between $T$=65K and $T$=85K. The bump grows from zero field to reach a maximum close to $H_{z}$=6Oe, for
a temperature close to $T$=80K. The sharp peak has a maximum at a temperature $T_{p}$ which is weakly field-dependent
(see upper inset to Fig.\ref{fig2}), and the full width at half-maximum FWHM broadens (linearly) with field (see lower
inset to Fig.\ref{fig2}).

\begin{figure}
\begin{center}
\includegraphics{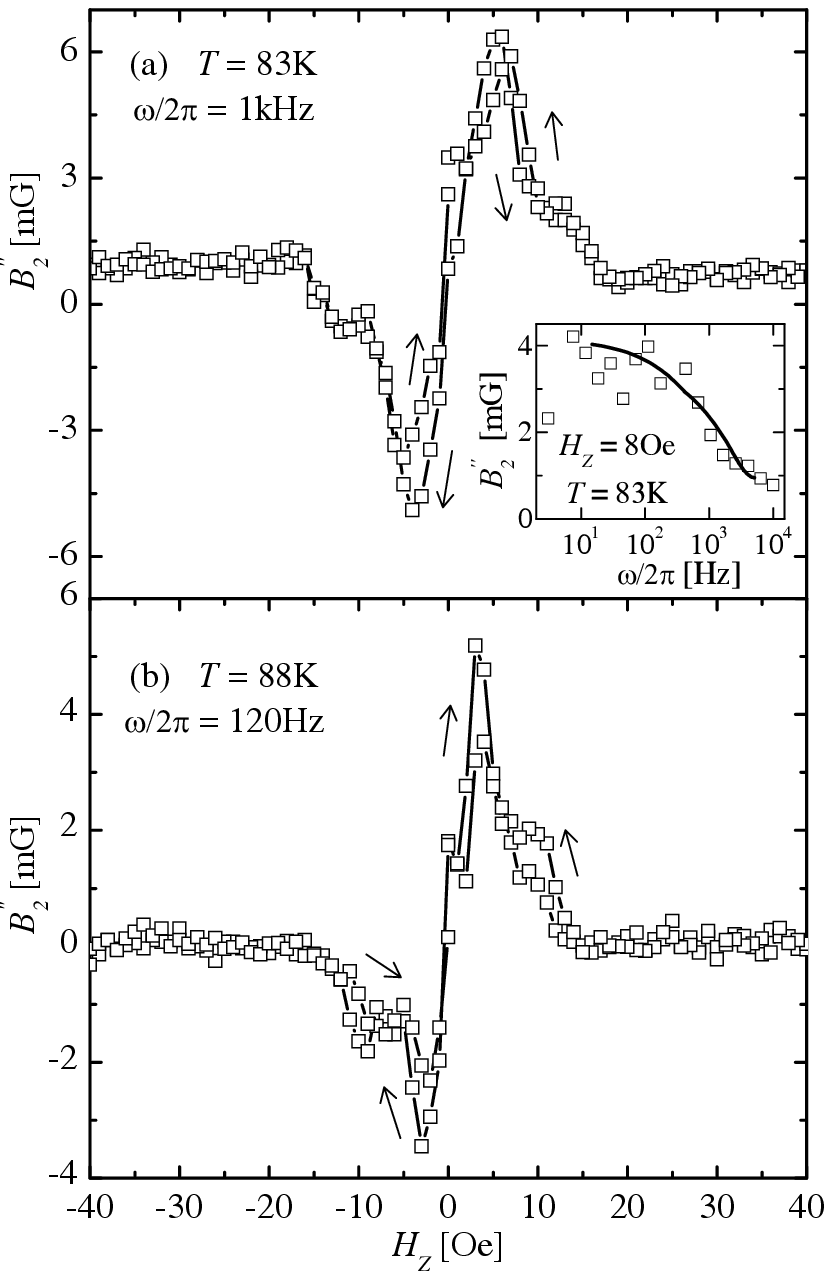}
\end{center}
\caption{\label{fig3} Quadrature component of the 2nd harmonic signal $B''_{2}$ measured with an excitation amplitude
$H^{AC}_{ab}$=2Oe, and (a) frequency $\omega/2\pi$=1kHz at $T$=83K and (b) $\omega/2\pi=120$Hz at $T$=88K as a function
of $H_{z}$ swept from positive to negative values starting at residual field. Arrows indicate field direction. Inset:
Frequency dependence of $B''_{2}$ at $T$=83K and for $H_{z}$=8Oe. The line is a guide for the eye.}
\end{figure}

Fig.\ref{fig3} establishes the sharp dependence of the response on PV density, as the field dependence of the two
features is explored by sweeping $H_{z}$ at particular temperatures, in the bump region and close to the peak maximum.
The dependence is similar for both features, showing an optimum pancake density where the response is maximal. For a
specific temperature and perpendicular field, we also find that the PV-JV interaction drops for frequencies above
$\approx$1kHz, as shown in the inset of Fig.\ref{fig3}. It is important to stress that here PVs are vibrating close to
their equilibrium lattice sites (whereas in a ratchet experiment when pancake vortices are dragged macroscopic
distances, high frequency interactions could be impeded still further \cite{Cole-ratchet}).

There are likely to be several mechanisms at play in these experiments including the temperature and field dependence
of the pancake lattice stiffness and the role of pinning. The bump feature is likely to be a direct measure of the
strength of the interaction between JVs and PVs, the interaction vanishing at zero field (no PVs), and being limited
when the PV repulsion dominates over the dragging interaction at higher fields. The coupling between the two vortex
species increases as the temperature is lowered and becomes arrested, at low temperature, when vortex pinning is so
strong as to prevent PV motion. Clues about the enhanced response just below $T_{c}$ comes from the field dependence of
the peak whose temperature $T_{p}$ drops at higher pancake densities. The $T_{p}$ vs $H_{z}$ curve  mimics the vortex
melting line observed in $Bi_{2}Sr_{2}CaCu_{2}O_{8}$ single crystals \cite{Zeldov-melting line} and clearly the
interaction between the two species of vortices is affected by the process of melting. We can tentatively associate
$T_{p}$ with the softening of the pancake lattice just before the lattice melts. Above the melting line the pancake
vortices behave in a liquid state and are free to move independently in response to the JV driving force. Below the
melting line the interaction strength between the two sublattices may be too weak for us to observe any signature of
interaction until the crystal is cooled further down in temperature. Interestingly two distinct peaks were previously
observed in the temperature behaviour of the microwave absorption in $Bi_{2}Sr_{2}CaCu_{2}O_{8}$ crystals and
attributed to two exotic collective modes of the Josephson plasma resonance associated with interaction with pancake
vortices \cite{Kakeya-Josephson plasma resonance}. However, as we do not observe any shift of the peak feature with
temperature when changing the frequency of the AC oscillation by two orders of magnitude, this strongly suggests that
our observations are unlikely related to a coherent collective plasma like resonance in the system.

\section{Conclusions}

In this paper we confirm the existence of DC lensing by monitoring the change in PV density induced by manipulation of
JVs. These observations stimulated us to use the highly sensitive second harmonic method to chart the strength of the
PV-JV interactions as a function of temperature and field. We have observed an unexpected and striking regime, very
close to the critical temperature where these interactions are enhanced. These results aid our understanding of how the
JVs might be used for lensing and also, by extension, for the manipulation of PVs in ratcheting experiments. These very
sensitive measurements of the dynamic interactions yield the optimum temperature and magnetic field conditions as well
as frequency range for the strongest coupling.

\ack 

We would like to thank Y. Yeshurun for stimulating discussions. The work was supported by  Leverhulme grant F/07 058/V
and T. Tamegai would like to acknowledge the Japanese Ministry of Education, Culture, Sports, Science, and Technology
for grant aid support.


\section*{References}

\begin{thebibliography}{13}

\bibitem{savelev-fluxonics devices}
Savel'ev S and Nori F 2002 Experimentally realizable devices for controlling the motion of magnetic flux quanta in
anisotropic superconductors \textit{Nature Mater.} \textbf{1} 179-184

\bibitem{Olson-pumps}
Olson C J, Reichhardt C, Janko B and Nori F 2001 Collective interaction-driven ratchet for transporting flux quanta
\textit{Phys. Rev. Lett.} \textbf{87} 177002

\bibitem{Wambaugh-pumps}
Wambaugh J F, Reichhardt C, Olson C J, Marchesoni F and Nori F 1999 Superconducting fluxon pumps and lenses
\textit{Phys. Rev. Lett.} \textbf{83} 5106-09

\bibitem{review}
Bending S J and Dodgson M J W 2005 Vortex chains in anisotropic superconductors \textit{J. Phys.: Condens. Matter}
\textbf{17}, R955-R993

\bibitem{Grigorenko-chain state}
Grigorenko A, Bending S J, Tamegai T, Ooi S and Henini M A 2001 A one-dimensional chain state of vortex matter
\textit{Nature} \textbf{414}, 728-731

\bibitem{Koshelev-tilted field}
Koshelev A E 1999 Crossing lattices, vortex chains, and angular dependence of melting line in layered superconductors
\textit{Phys. Rev. Lett.} \textbf{83}, 187-190

\bibitem{Cole-ratchet}
Cole D, Bending S J, Savel'ev S, Grigorenko A, Tamegai T and Nori F 2006 Ratchet without spatial asymmetry for
controlling themotion of magnetic flux quanta using time-asymmetric drives \textit{Nature Mater.} \textbf{5}, 305-311

\bibitem{Perkins-dynamics interaction}
Perkins G K, Caplin A D and Cohen L F 2005 Dynamic interactions between pancake vortex stacks and Josephson vortices in
$Bi_{2}Sr_{2}CaCu_{2}O_{8}$ single crystals: relaxation and ratchets \textit{Supercond. Sci. Technol.} \textbf{18},
1290-93

\bibitem{Tamegai-samples}
Ooi S, Shibauchi T and Tamegai T 1998 Evolution of vortex phase diagram with oxygen-doping in
$Bi_{2}Sr_{2}CaCu_{2}O_{8+y}$ single crystals \textit{Physica C} \textbf{302}, 339-345

\bibitem{Cole-lensing}
Cole D, Bending S J, Savel'ev S, Tamegai T and Nori F 2006 Manipulation of magnetic-flux landscapes in superconducting
$Bi_{2}Sr_{2}CaCu_{2}O_{8+\delta}$ crystals \textit{Europhys. Lett.} \textbf{76}, 1151-57

\bibitem{Tamegai-lensing}
Tamegai T, Chiku H, Aoki H and Tokunaga M 2006 Visualization and control of vortex chains in highly anisotropic
superconductors \textit{Physica C} \textbf{437}, 314-318

\bibitem{Zeldov-melting line}
Khaykovich B, Zeldov E, Majer D, Li T W, Kes P H and Konczykowski M 1996 Vortex-Lattice Phase Transitions in
$Bi_{2}Sr_{2}CaCu_{2}O_{8}$ Crystals with Different Oxygen Stoichiometry \textit{Phys. Rev. Lett.} \textbf{76}, 2555-58

\bibitem{Kakeya-Josephson plasma resonance}
Kakeya I, Wada T, Nakamura R and Kadowaki K 2005 Two phase collective modes in a Josephson vortex lattice in the
intrinsic Josephson junction $Bi_{2}Sr_{2}CaCu_{2}O_{8+\delta}$ \textit{Phys. Rev. B} \textbf{72}, 014540

\end{thebibliography}
\end{document}